\documentclass[reprint,aps,pra,longbibliography,superscriptaddress]{revtex4-2}

\usepackage{amssymb, amsmath}
\usepackage[english]{babel}
\usepackage{bm}
\usepackage{graphicx}
\usepackage[utf8]{inputenc}
\usepackage{bbold}
\usepackage{hyperref}

\hypersetup{colorlinks=true, citecolor=blue, urlcolor=teal, linkcolor=blue}

\usepackage[dvipsnames]{xcolor}

\newcommand{\red}[1]{#1}

\mathchardef\myminus="2D

\newcommand\mydots{...\,}

\usepackage{letltxmacro}
\LetLtxMacro{\ORIGselectlanguage}{\selectlanguage}
\makeatletter
\DeclareRobustCommand{\selectlanguage}[1]{%
  \@ifundefined{alias@\string#1}
    {\ORIGselectlanguage{#1}}
    {\begingroup\edef\x{\endgroup
       \noexpand\ORIGselectlanguage{\@nameuse{alias@#1}}}\x}%
}
\newcommand{\definelanguagealias}[2]{%
  \@namedef{alias@#1}{#2}%
}
\newcommand{\opvec}[1]{\hat{\mathbf{#1}}}
\newcommand{\qnm}[2]{\tilde{\mathbf{#1}}_{#2}}

\makeatletter
\def\maketitle{
\@author@finish
\title@column\titleblock@produce
\suppressfloats[t]}
\makeatother 

\definelanguagealias{en}{english}
\definelanguagealias{EN}{english}
\definelanguagealias{eng}{english}
\definelanguagealias{de}{ngerman}

\begin{document}
\author{Robert Meiners Fuchs}
\email[]{r.fuchs.1@tu-berlin.de}
\affiliation{Institut für Physik und Astronomie, Nichtlineare Optik und
Quantenelektronik, Technische Universität Berlin, Hardenbergstr. 36, EW 7-1, 10623
Berlin, Germany}

\author{Juanjuan Ren}
\affiliation{Department of Physics, Engineering Physics, and Astronomy,
Queen’s University, Kingston, Ontario K7L 3N6, Canada}

\author{Stephen Hughes}
\affiliation{Department of Physics, Engineering Physics, and Astronomy,
Queen’s University, Kingston, Ontario K7L 3N6, Canada}

\author{Marten Richter}
\email[]{marten.richter@tu-berlin.de}
\affiliation{Institut für Physik und Astronomie, Nichtlineare Optik und
Quantenelektronik, Technische Universität Berlin, Hardenbergstr. 36, EW 7-1, 10623
Berlin, Germany}

\title{Green's function expansion for multiple coupled optical resonators with finite retardation using quasinormal modes}



\begin{abstract}
The electromagnetic Green's function is a crucial ingredient for the theoretical study of modern photonic quantum devices, but is often difficult or even impossible to calculate directly. We present a numerically efficient framework for calculating the scattered electromagnetic Green's function of a multi-cavity system with spatially separated open cavities (with arbitrary shape, dispersion and loss) and finite retardation times. The framework is based on a Dyson scattering equation that enables the construction of the Green's function from the quasinormal modes of the individual resonators within a few-mode approximation and a finite number of iteration steps without requiring nested integrals. The approach shows excellent agreement with the full numerical Green's function for the example of two coupled dipoles located in the gaps of two metal dimers serving as quasinormal mode cavities, and is easily extended to arbitrarily large separations and multiple cavities.
\end{abstract}


\date{\today}
\maketitle
\textit{Introduction.---}Open optical cavity resonators coupled to quantum emitters
(such as atoms, molecules, and quantum dots) are a fundamental building block for many modern quantum technologies such as nanolasers \cite{an1994microlaser, pscherer2021single, rivero2021non}, quantum sensors \cite{posani2006nanoscale, lee2021quantum}, quantum computers \cite{pellizzari1995decoherence, blais2004cavity, benito2019optimized, borjans2020resonant}, and quantum networks \cite{cirac1997quantum, pellizzari1997quantum, duan2001long, pichler2016photonic}.
For a theoretical treatment of such systems, knowledge of the full electromagnetic Green's function is of great importance and a {\it requirement} to many quantum optics formalisms. For example, it is used to calculate the Purcell enhancement of optical cavities \cite{vogel2006quantum, sauvan2013theory, franke2019quantization, ren2021quasinormal}, in the quantization of the electromagnetic field \cite{ho1998second, dung1998three, suttorp2004field, philbin2010canonical, franke2019quantization, fuchs2024quantization}, and in the derivation of coupling elements between photons and quantum emitters \cite{franke2020quantized, medina2021few}. Therefore, an efficient scheme for obtaining a good approximation of the full Green's function is central to many theoretical treatments of light-matter coupling in such systems. 

One approach to open cavity systems is through quasinormal modes (QNMs), which are solutions to the source-free Helmholtz equation under open boundary conditions \cite{leung1994completeness, muljarov2011brillouin, kristensen2012generalized, sauvan2013theory}. The QNMs have complex eigenfrequencies with a negative imaginary part, so that the temporal decay is inherent to the mode, making the QNMs the {\it natural modes} of open cavity systems with material losses. \red{Quasinormal mode theories have been used to study a variety of setups in recent works \cite{sauvan2013theory, weiss2016from, yan2020shape, primo2020quasinormal, gustin2025what, muljarov2026rigorous}, and 
play a crucial role in rigorous open cavity quantum optics \cite{ho1998second, franke2019quantization, fuchs2024quantization}.}

For positions within (or close to) the resonator, an expansion in terms of a few dominant QNMs yields a very good approximation of the full Green's function \cite{leung1994completeness, muljarov2011brillouin, ge2014quasinormal, franke2019quantization, kristensen2020modeling}. For positions far away from the resonator, however, the QNMs diverge due to their complex eigenfrequencies. Even for properly normalized modes, a large number of QNMs is usually necessary for a good approximation of the Green's function \cite{abdelrahman2018completeness, sztranyovszky2025extending}. This makes the approach impractical for many quantum optics or quantum dynamics applications, where the Hilbert space scales exponentially with the number of modes.

For single resonators, the QNMs outside the cavity region can be replaced with {\it regularized frequency-dependent fields} \cite{ge2014quasinormal, ren2020near, franke2020quantized}, which can be calculated efficiently using a near-field-to-far-field transformation together with a pole approximation \cite{ren2020near}. For many applications, this approach makes a few-QNM expansion sufficient, even for positions far away from the resonator.

However, many modern quantum devices consist of multiple, spatially separated and interacting cavities. A calculation of the exact QNMs of such coupled structures is often not feasible. It is instead desirable to express the multi-cavity Green's function in terms of the single-cavity QNMs. For example, in the \textit{coupled quasinormal mode theory} (CQT) \cite{PhysRevB.102.045430,ren2021quasinormal, ren2022connecting}, symmetrized eigenfrequencies and hybridized QNMs are derived from single-cavity properties. However, CQT relies on divergent QNMs, and is therefore {\it not accurate} for large separations.

In this Letter, we introduce a powerful and accurate Dyson equation approach to obtain the multi-cavity scattered Green's function \red{for systems with large spatial separations and non-negligible retardation} from only single-cavity QNMs, within a few-mode approximation. We use regularized QNM fields outside the individual resonators to characterize the intercavity scattering, \red{so that retardation effects are included}. Multi-cavity scattering naturally decomposes into products of two-cavity scattering processes within the QNM expansion, {\it avoiding nested integrals}. \red{This approach allows for a few-mode expansion of the multi-cavity Green's function, and thus the implementation in quantum optics applications.} We calculate the coupling between two dipole emitters located in the gaps of metal dimers serving as QNM cavities, and compare the QNM expansion with the full numerical Green's function to excellent agreement.

\textit{Coupled-cavity system.---}We consider a system of \(N\) spatially separated cavities or plasmonic nanoparticles. The full permittivity,
satisfying causality, reads
\begin{align} \label{eq:fullperm}
    \epsilon(\mathbf{r},\omega) = \epsilon_{\rm back}(\mathbf{r},\omega) +\sum_{i=1}^N V_i(\mathbf{r},\omega),
\end{align}
where \(\epsilon_{\rm back}(\mathbf{r},\omega)\) is the permittivity of the background medium, \(V_i(\mathbf{r},\omega) = \chi_{\mathcal{V}_i}(\mathbf{r})\big[\epsilon(\mathbf{r},\omega)-\epsilon_{\rm back}(\mathbf{r},\omega)\big]\)
is the perturbation of the permittivity due to the presence of the cavity, with \(\mathcal{V}_i\) the cavity volume, and the function \(\chi_{\mathcal{V}_i}(\mathbf{r})\) is unity for \(\mathbf{r}\in\mathcal{V}_i\) and zero elsewhere. \red{We make no assumption about the background permittivity \(\epsilon_{\rm back}\) except those necessary to define QNMs, i.e., that there is a discontinuity between the permittivity of the cavity medium and \(\epsilon_{\rm back}\), and that there is no backscattering of waves from the far field \cite{ching1998quasinormal}.}

The full electromagnetic Green's tensor (or dyad) is the solution to the Helmholtz equation
\begin{align} \label{eq:greenhelm}
    \Big[\nabla\times\nabla\times-\frac{\omega^2}{c^2}\epsilon(\mathbf{r},\omega)\Big]\mathbf{G}(\mathbf{r},\mathbf{r}',\omega)=\frac{\omega^2}{c^2}\mathbb{1}\delta(\mathbf{r}-\mathbf{r}'),
\end{align}
under appropriate boundary conditions\red{, which is sufficient for defining the Green's function mathematically}. For cavities in vacuum, as considered in this paper, this is the Silver-Müller radiation condition \cite{muller1948grundzuge, silver1984microwave}.

\textit{Single-cavity QNMs.---}We define the single-cavity permittivity 
\(\epsilon_i(\mathbf{r},\omega) = \epsilon_{\rm back}(\mathbf{r},\omega) +V_i(\mathbf{r},\omega)\),
where the permittivity is set to the background value everywhere except at the \(i\)-th cavity. The QNMs \(\qnm{f}{i_{\mu}}(\mathbf{r})\) of the \(i\)-th cavity solve the source-free Helmholtz equation for this (single-cavity) permittivity \cite{leung1994completeness, muljarov2011brillouin, kristensen2012generalized, sauvan2013theory, muljarov2026rigorous},
\begin{align}
    \nabla\times\nabla\times\qnm{f}{i_{\mu}}(\mathbf{r})-\frac{\tilde{\omega}_{i_{\mu}}^2}{c^2}\epsilon_i(\mathbf{r},\tilde{\omega}_{i_{\mu}})\qnm{f}{i_{\mu}}(\mathbf{r}) = 0,
\end{align}
under open boundary conditions, such as the Silver-Müller radiation condition for cavities in a homogeneous background medium \cite{kristensen2020modeling, franke2019quantization}, or waveguide radiation conditions for waveguide-coupled cavities \cite{kristensen2014calculation}. 
Due to the outgoing boundary conditions and complex permittivity, QNMs have complex eigenfrequencies with negative imaginary part: \(\tilde{\omega}_{i_{\mu}} = \omega_{i_{\mu}}-i\gamma_{i_{\mu}},\, \gamma_{i_{\mu}}>0\).

We assume that, near the resonator, the single-cavity Green's function can be expanded as usual in terms of the QNMs \(\qnm{f}{i_{\mu}}(\mathbf{r})\), so that \cite{leung1994completeness, muljarov2011brillouin, ge2014quasinormal, franke2019quantization, kristensen2020modeling}
\begin{align}\label{eq:singlegreen}
    &\mathbf{G}_i(\mathbf{r},\mathbf{r}',\omega)\big|_{\mathbf{r},\mathbf{r}'\in \mathcal{V}_i} = \sum_{\mu}A_{i_{\mu}}(\omega)\qnm{f}{i_{\mu}}(\mathbf{r})\qnm{f}{i_{\mu}}(\mathbf{r}'),
\end{align}
where \(A_{i_{\mu}}(\omega) = \omega/[2(\tilde{\omega}_{i_{\mu}}-\omega)]\).

For positions far away from the resonator, the QNMs spatially diverge due to the complex eigenfrequencies \cite{muljarov2011brillouin, sauvan2013theory, kristensen_modes_2014}. Even if properly normalized, a large number of QNMs is generally required for an accurate expansion of the Green's function \cite{abdelrahman2018completeness, sztranyovszky2025extending}. A common alternative replaces the QNMs outside the cavity volume with non-divergent frequency-dependent regularized QNM fields \(\qnm{F}{i_{\mu}}(\mathbf{r},\omega)\), which are obtained from a Dyson equation \cite{ge2014quasinormal}: 
\begin{align}
    \qnm{F}{i_{\mu}}(\mathbf{r},\omega) = \int_{\mathcal{V}_i}\mathrm{d}^3r' \Delta\epsilon(\mathbf{r}',\omega)\mathbf{G}_{\rm back}(\mathbf{r},\mathbf{r}',\omega)\cdot\qnm{f}{i_{\mu}}(\mathbf{r}'),
\end{align}
where \(\Delta\epsilon(\mathbf{r},\omega) = \epsilon(\mathbf{r},\omega)-\epsilon_{\rm back}(\mathbf{r},\omega)\), and \(\mathbf{G}_{\rm back}(\mathbf{r},\mathbf{r}',\omega)\) is the Green's function of the background medium, which solves the Helmholtz equation~\eqref{eq:greenhelm} with the background permittivity \(\epsilon_{\rm back}(\mathbf{r},\omega)\). For a more efficient calculation, regularized QNMs can be obtained from integrals over the cavity surface only, using the field-equivalence principle \cite{franke2023impact} or a near-field-to-far-field 
transformation~\cite{ren2020near}.

\begin{figure}
    \centering
    \includegraphics[width=0.8\linewidth]{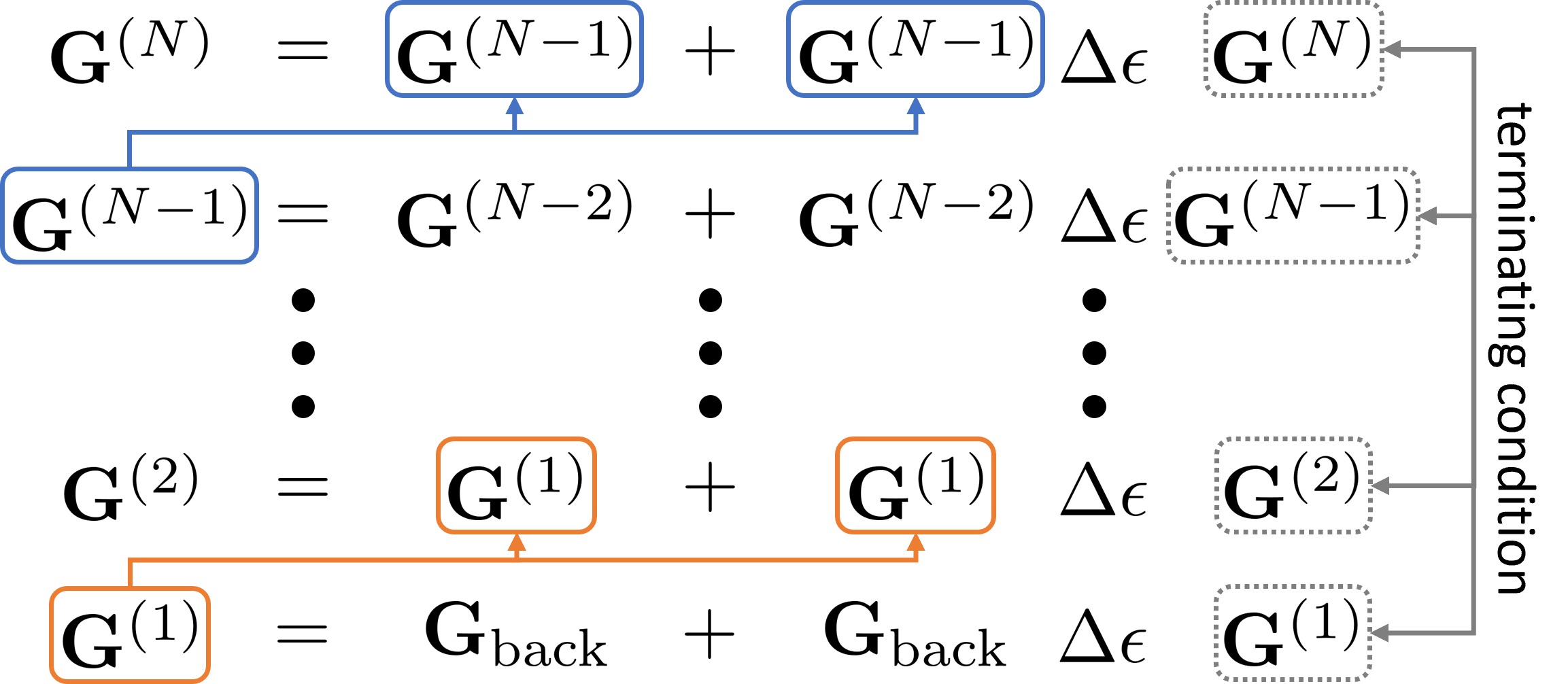}
    \caption{Sketch of the framework for obtaining the \(N\)-cavity Green's function via a set of \(N\) scattering equations. Starting from the single-cavity Green's function \(\mathbf{G}^{(1)}\), we iteratively add more cavities with \(\mathbf{G}^{(n-1)}\) from the previous step serving as an input for the scattering equation for the Green's function \(\mathbf{G}^{(n)}\) [cf.~Eq.~\eqref{eq:dyson}]. In each step, the recursive Dyson equation is terminated using the condition from Eq.~\eqref{eq:termcond}.}
    \label{fig:framework}
\end{figure}
\textit{Multi-cavity Green's 
function expansion.---}We wish  to construct the Green's function for \(N\) cavities from the single-cavity QNMs. We do this via a set of scattering equations where, starting from the single-cavity Green's function, we iteratively add more cavities until all \(N\) cavities are included. We sketch this framework in Fig.~\ref{fig:framework}. 

Let \(\mathbf{G}^{(n)}(\mathbf{r},\mathbf{r}',\omega)\) be the Green's function for \(n\) cavities which solves the Helmholtz equation~\eqref{eq:greenhelm} for 
\begin{align}\label{eq:nperm}
    \epsilon^{(n)}(\mathbf{r},\omega) = \epsilon_{\rm back}(\mathbf{r},\omega) +\sum_{i=1}^n V_i(\mathbf{r},\omega).
\end{align}

Now, \(\mathbf{G}^{(n)}(\mathbf{r},\mathbf{r}',\omega)\) can be obtained from the Green's function \(\mathbf{G}^{(n-1)}(\mathbf{r},\mathbf{r}',\omega)\) for \(n-1\) cavities via the scattering equation [see \red{S2 A in} Ref.~\onlinecite{Supp} for the derivation] 
\begin{align}\label{eq:dyson}
    &\mathbf{G}^{(n)}(\mathbf{r},\mathbf{r}',\omega) = \mathbf{G}^{(n-1)}(\mathbf{r},\mathbf{r}',\omega)\nonumber\\
    &\;+\int_{\mathcal{V}_n}\mathrm{d}^3s \Delta\epsilon(\mathbf{s},\omega)\mathbf{G}^{(n-1)}(\mathbf{r},\mathbf{s},\omega)\cdot\mathbf{G}^{(n)}(\mathbf{s},\mathbf{r}',\omega),
\end{align}
where \(\mathcal{V}_n\) is the volume of the \(n\)-th cavity. For \(n=1\) and using \(\mathbf{G}^{(0)}\equiv\mathbf{G}_{\rm back}\), we recover the Dyson equation for the single-cavity case from Ref.~\onlinecite{ge2014quasinormal}.

For any fixed \(n=k\), Eq.~\eqref{eq:dyson} is recursive, since \(\mathbf{G}^{(k)}\) appears also on the right-hand side (RHS) of Eq.~\eqref{eq:dyson}. \red{While a perturbative treatment of Eq.~\eqref{eq:dyson} is generally possible (cf.~Sec.~S2 C in \cite{Supp}),} this \(\mathbf{G}^{(k)}\) on the RHS 
{\it only} depends on positions within the same cavity \footnote{For any fixed $n=k$, either (i)  \(\mathbf{r}'\in \mathcal{V}_{k}\), or (ii) \red{if \(\mathbf{r}\in \mathcal{V}_{k}\), case (i) is recovered} by using the transpose of Eq.~\eqref{eq:dyson}, or (iii) if neither $\mathbf{r}'$ nor $\mathbf{r}$ are in  $\mathcal{V}_{k}$
inserting the transpose of Eq.~\eqref{eq:dyson} on the RHS of Eq.~\eqref{eq:dyson} again recovers case (i).}.
Consequently, we can terminate the Dyson equation at low order for any $k$ by replacing $\mathbf{G}^{(k)}$ at the RHS with the single-cavity QNM Green's function (\(1\leq i \leq k\))
\begin{align}\label{eq:termcond}
    \mathbf{G}^{(k)}(\mathbf{r},\mathbf{r}',\omega)\big|_{\mathbf{r},\mathbf{r}'\in \mathcal{V}_i} \approx \mathbf{G}_i(\mathbf{r},\mathbf{r}',\omega), 
\end{align}
i.e., we assume that the Green's function for positions inside the cavity can be approximated by the single-cavity Green's function of that cavity [cf.~Eq.~\eqref{eq:singlegreen}]. This assumption holds well for many applications, such as 
3D cavities in homogeneous media \red{(where the influence of other cavities scales as \(1/R\)) with separations on the order of the QNM wavelength or more (see Fig.~\ref{fig:Gcomp_d740} as well as Figs.~S2 and S3 in Ref.~\onlinecite{Supp}). For cases where the assumption breaks down, the perturbative approach (see S2 C in \cite{Supp}) or the established CQT are employed.}

As a consequence, the full Green's function for \(N\) cavities, \(\mathbf{G}(\mathbf{r},\mathbf{r}',\omega) = \mathbf{G}^{(N)}(\mathbf{r},\mathbf{r}',\omega)\) is obtained iteratively from a set of \(N\) scattering equations~\eqref{eq:dyson} for \(n=1,\mydots,N\) (cf.~Fig.~\ref{fig:framework}). 

As we show for three coupled cavities in S2 D of Ref.~\onlinecite{Supp} the ordering of the cavities \(1,\mydots,N\) should be performed such that the Dyson equation~\eqref{eq:dyson} terminates [i.e., the case from Eq.~\eqref{eq:termcond} is obtained on the RHS] at low order for each step in the iteration.

This formalism yields the scattered Green's function for an arbitrary number of cavities within a finite number of iterations. In addition, the QNM expansion factorizes the Green's function [cf.~Eq.~\eqref{eq:singlegreen}], so that multi-cavity scattering processes naturally decompose into products of two-cavity scattering processes (see S2 D in Ref.~\onlinecite{Supp} for an example of three-cavity scattering), thus avoiding numerically demanding nested integrals. 

For a practical and efficient numerical calculation, the volume integral representation from Eq.~\eqref{eq:dyson} can be converted into surface integrals\red{. Towards, this, we note that from Eq.~\eqref{eq:nperm}, it follows that \(\Delta\epsilon(\mathbf{s},\omega)\big|_{\mathbf{s}\in\mathcal{V}_n} = \epsilon^{(n)}(\mathbf{s},\omega)-\epsilon^{(n-1)}(\mathbf{s},\omega)\) in Eq.~\eqref{eq:dyson}. Thus, we employ the Helmholtz equation~\eqref{eq:greenhelm} with \(\epsilon^{(n)}\) for \(\mathbf{G}^{(n)}\) and with \(\epsilon^{(n-1)}\) for \(\mathbf{G}^{(n-1)}\) in Eq.~\eqref{eq:dyson}, to find via the second Green's theorem} (omitting the \(\omega\)-dependence in the Green's functions for a brief notation, \red{see Sec.~S2 B in Ref.~\onlinecite{Supp} for details}):
\begin{align}\label{eq:dysonsurface}
    &\frac{\omega^2}{c^2}\mathbf{G}^{(n)}(\mathbf{r},\mathbf{r}')\Big|_{\mathbf{r}\notin\mathcal{V}_n}= \chi_{\overline{\mathcal{V}}_n}(\mathbf{r}')\frac{\omega^2}{c^2}\mathbf{G}^{(n-1)}(\mathbf{r},\mathbf{r}')\nonumber\\
    &+\;\oint_{\mathcal{S}_n}\mathrm{d}A_s \Big\{\Big[\nabla_s\times\mathbf{G}^{(n-1)}(\mathbf{s},\mathbf{r})\Big]^T\cdot\Big[\opvec{n}_s\times\mathbf{G}^{(n)}(\mathbf{s},\mathbf{r}')\Big]\nonumber\\
    &\qquad\qquad-\Big[\opvec{n}_s\times\mathbf{G}^{(n-1)}(\mathbf{s},\mathbf{r})\Big]^T\cdot\Big[\nabla_s\times\mathbf{G}^{(n)}(\mathbf{s},\mathbf{r}')\Big]\Big\},
\end{align}
where \(\overline{\mathcal{V}}_n\) is the complement of the cavity volume \(\mathcal{V}_n\), and \(\mathcal{S}_n\) is a closed surface around \(\mathcal{V}_n\) with the outward-pointing surface normal vector \(\opvec{n}_s\). 

In principle, the volume integral in Eq.~\eqref{eq:dyson} and the surface integral in Eq.~\eqref{eq:dysonsurface} should yield the same result for the Green's function. However, since different parts of the QNMs dominate inside the cavity volume and on the surface, the two representations only fully agree when all QNMs are included \cite{franke2023impact}. Hence, one representation should be used consistently, e.g., if the regularized QNM fields \(\qnm{F}{i_{\mu}}(\mathbf{r},\omega)\) are calculated using surface integrals (as we do below), the representation from Eq.~\eqref{eq:dysonsurface} should be used. We note that the use of surface integrals can also help circumvent problems arising in the context of perturbation theories with volume integrals due to the discontinuity of the electric field at the boundary between media \cite{johnson2002perturbation, johnson2005roughness, patterson2010interplay, yan2020shape}.

\textit{Coupled metal 
dimers.---}For an illustrative practical example, we consider two metal dimers with volumes \(\mathcal{V}_1\) and \(\mathcal{V}_2\) in vacuum as QNM cavities \cite{Supp}. We assume that each dimer is dominated by a single QNM and described by the Drude model
\begin{align}\label{eq:Drude}
    \epsilon_j(\omega)=1-\frac{\omega_{\rm p}^{2}}{\omega^{2}+i\omega\gamma_{{\rm p}j}},
 \end{align}
where, \(j=1,2\), and \(\hbar\omega_{\rm p} = 8.2934\,{\rm eV}\)\red{, \(\hbar\gamma_{\rm p1}=0.0928\,{\rm eV}\) are the plasma frequency and damping rate for gold, while dimer 2 has a reduced damping rate of \(\hbar\gamma_{\rm p2}=0.3\hbar\gamma_{\rm p1}\) to obtain a higher-quality plasmon mode.} 
The QNM of dimer \(1\) has the eigenfrequency \(\hbar\tilde{\omega}_1=(1.6904-i0.0652)\,{\rm eV}\), corresponding to a wavelength of \(\lambda_1 \approx 733.46\,{\rm nm}\). For the QNM of dimer \(2\), we obtain \(\hbar\tilde{\omega}_2=(1.6482-i0.0388)\,{\rm eV}\) and \(\lambda_2 \approx 752.24\,{\rm nm}\).

\begin{figure}
    \centering
    \includegraphics[width=0.8\linewidth]{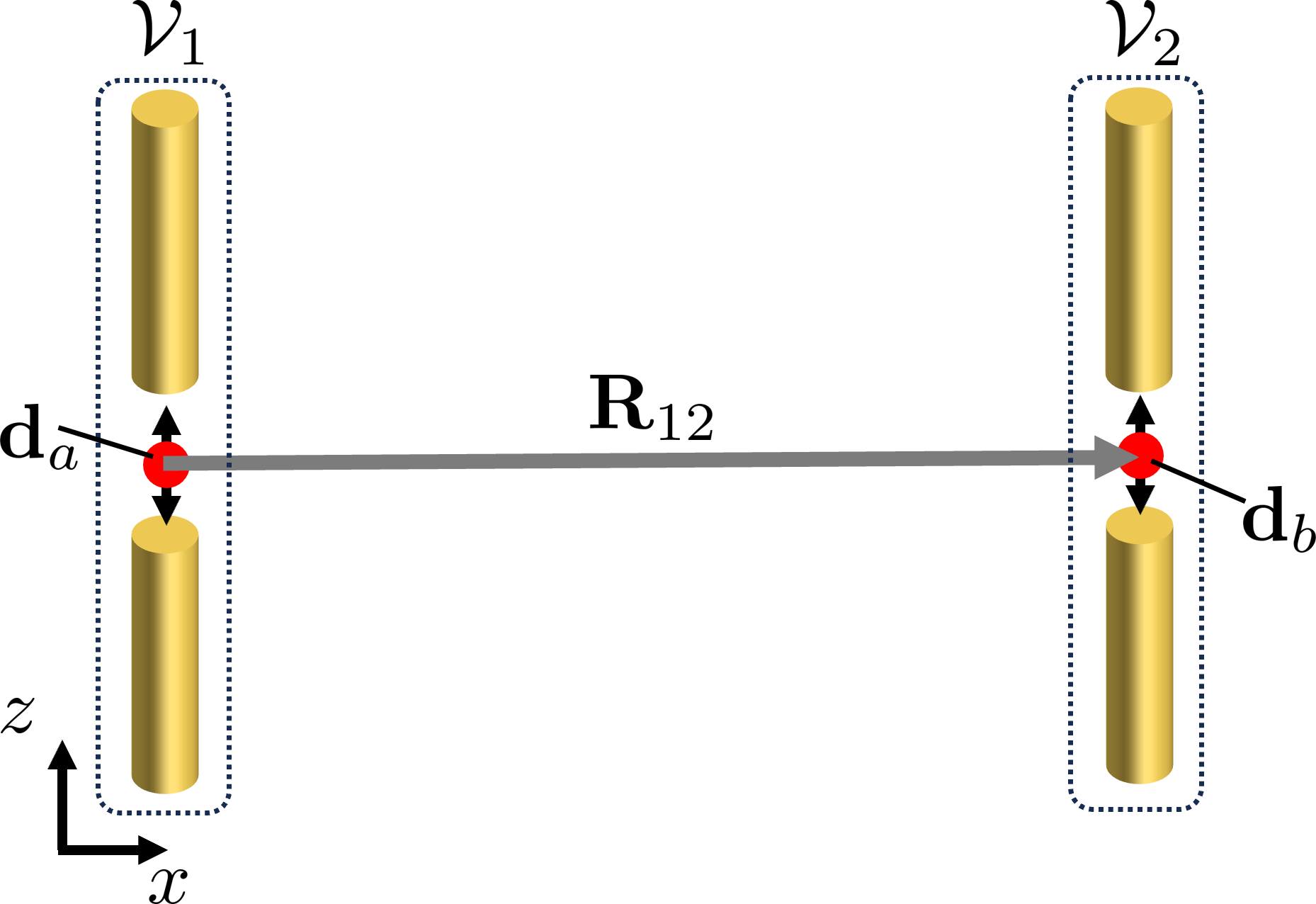}
    \caption{Sketch of two metal dimers serving as QNM cavities. The dimers are described by a Drude permittivity [cf.~Eq.~\eqref{eq:Drude}], and separated by he center-to-center distance \(R_{12}\). Two dipole emitters \(\mathbf{r}_a\in \mathcal{V}_1\) and \(\mathbf{r}_b\in \mathcal{V}_2\) are placed in the dimer gaps.}
    \label{fig:sketch}
\end{figure}

As shown in Fig.~\ref{fig:sketch}, we place two identical \(z\)-polarized dipoles \(\mathbf{d}_a=\mathbf{d}_b= d\opvec{e}_z\) in the dimer gaps (\(\mathbf{r}_a\in \mathcal{V}_1, \mathbf{r}_b \in \mathcal{V}_2\)). We use the coupling between the dipoles,
\begin{align}\label{eq:gba}
    g_{ba}(\omega) = \frac{1}{\epsilon_0}\mathbf{d}_b\cdot\mathbf{G}(\mathbf{r}_b,\mathbf{r}_a,\omega)\cdot\mathbf{d}_a,
\end{align}
as an indicator to compare the QNM expansion from Eq.~\eqref{eq:dysonsurface} to the full Green's function, \red{which is obtained numerically from the scattered electric field of a point dipole source (see S1 in \cite{Supp})}.

\red{To obtain \(\mathbf{G}(\mathbf{r}_b,\mathbf{r}_a,\omega)\) from Eq.~\eqref{eq:dysonsurface}, we use \(\mathcal{S}_n \to \mathcal{S}_1\) so that \(\mathbf{s},\mathbf{r}_a\in \mathcal{V}_1\), allowing us to apply the termination condition from Eq.~\eqref{eq:termcond}. We also employ the single-cavity QNM expansion \cite{franke2020quantized}
\begin{align}\label{eq:singlegreenoutside}
    \mathbf{G}^{(n-1)}(\mathbf{s},\mathbf{r}_b,\omega) &\to \mathbf{G}_2(\mathbf{s},\mathbf{r}_b,\omega)\big|_{\mathbf{s}\notin \mathcal{V}_2,\mathbf{r}_b\in\mathcal{V}_2}\nonumber\\
    & = A_2(\omega)\qnm{F}{2}(\mathbf{s},\omega)\qnm{f}{2}(\mathbf{r}_b).
\end{align}

Thus, we arrive at}
\begin{align}\label{eq:QNMexp}
    &\mathbf{G}(\mathbf{r}_b,\mathbf{r}_a,\omega)\big|_{\mathbf{r}_b \in \mathcal{V}_2,\mathbf{r}_a\in \mathcal{V}_1}=B_{21}(\omega)\qnm{f}{2}(\mathbf{r}_b)\qnm{f}{1}(\mathbf{r}_a),
\end{align}
where 
\begin{align}\label{eq:Bsurface}
    &B_{21}(\omega) = \frac{c^2}{\omega^2}A_2(\omega)A_1(\omega)\nonumber\\
    &\quad\qquad\quad\times\oint_{\mathcal{S}_1}\mathrm{d}A_s \Big\{\Big[\nabla_s\times\qnm{F}{2}(\mathbf{s},\omega)\Big]\cdot\Big[\opvec{n}_s\times\qnm{f}{1}(\mathbf{s})\Big]\nonumber\\
    &\qquad\qquad\qquad-\Big[\opvec{n}_s\times\qnm{F}{2}(\mathbf{s},\omega)\Big]\cdot\Big[\nabla_s\times\qnm{f}{1}(\mathbf{s})\Big]\Big\}.
\end{align}
Note that \(B_{21}(\omega) = B_{12}(\omega)\) holds, preserving \(\mathbf{G}(\mathbf{r}_a,\mathbf{r}_b,\omega) = [\mathbf{G}(\mathbf{r}_b,\mathbf{r}_a,\omega)]^T\) \red{for reciprocal media}.

For an efficient calculation of the expansion from Eq.~\eqref{eq:QNMexp}, we rewrite the regularized QNM field \(\qnm{F}{2}(\mathbf{s},\omega)\big|_{\mathbf{s}\in\mathcal{V}_1} = \qnm{F}{2}'(\mathbf{s},\omega)\mathrm{e}^{i\omega R_{21}/c}\), where \( \qnm{F}{2}'(\mathbf{s},\omega)\) is a slow-rotating envelope function, and \(R_{21}= R_{12} = 2020\, {\rm nm}\) is the center-to-center distance between the dimers \cite{fuchs2024quantization}. The envelope function \( \qnm{F}{2}'(\mathbf{s},\omega)\) is treated in a pole approximation \(\omega\to\omega_2\) near the poles contained in \(A_2(\omega), A_1(\omega)\), and we arrive at:
\begin{align}\label{eq:Bapprox}
    B_{21}(\omega)\approx N_{21} \frac{c^2}{\omega^2} A_2(\omega)A_1(\omega)\mathrm{e}^{i\omega R_{21}/c},
\end{align}
\red{with
\begin{align}\label{eq:N21}
    N_{21} &= \oint_{\mathcal{S}_1}\mathrm{d}A_s \Big\{\Big[\nabla_s\times\qnm{F}{2}'(\mathbf{s},\omega_2)\Big]\cdot\Big[\opvec{n}_s\times\qnm{f}{1}(\mathbf{s})\Big]\nonumber\\
    &\qquad\qquad-\Big[\opvec{n}_s\times\qnm{F}{2}'(\mathbf{s},\omega_2)\Big]\cdot\Big[\nabla_s\times\qnm{f}{1}(\mathbf{s})\Big]\Big\}.
\end{align} 
}
In Fig.~\ref{fig:Gcomp_d2000}, we show the imaginary and real parts of \(g_{ba}(\omega)\), calculated using the full numerical Green's function and the QNM approximation from Eq.~\eqref{eq:QNMexp} together with Eq.~\eqref{eq:Bapprox} \red{[\(N_{21} = (2.0773-i0.0657)\cdot 10^{-7}\,{\rm nm}^{-2}\)]}, and observe {\it excellent agreement}. We stress that {\it we do not  use any fitting parameters in the calculations}. 
The frequency-dependent phase \(\mathrm{e}^{i\omega R_{12}/c}\) from Eq.~\eqref{eq:Bapprox} is crucial for cases with significant separation. We show in Fig.~\ref{fig:Gcomp_d2000} the case where the exponential is included in the pole approximation (\(G^{\rm QNM}_{\rm pole}\)),
\begin{align}\label{eq:Bpole}
    B_{21}(\omega)\big|^{\rm pole}\approx N_{21} \frac{c^2}{\omega^2} A_2(\omega)A_1(\omega)\mathrm{e}^{i\omega_2 R_{21}/c},
\end{align}
which does not match the full numerical results, since it does not fully account for retardation effects.

\begin{figure}
    \centering
    \includegraphics[width=0.9\linewidth]{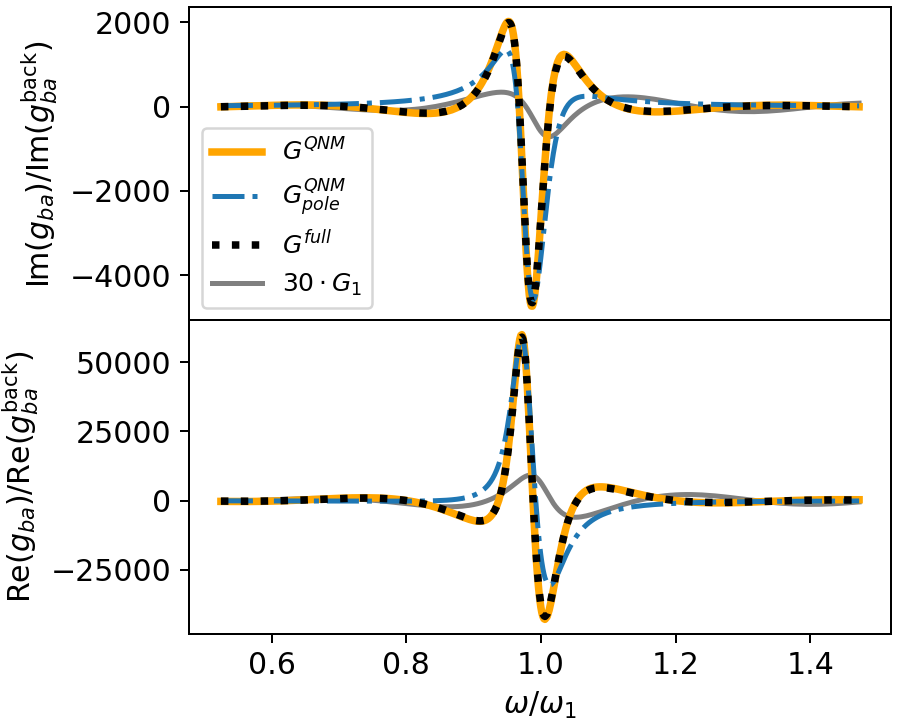}
    \caption{Normalized coupling \(g_{ba}(\omega)\) 
    [cf.~Eq.~\eqref{eq:gba}] between the two dipoles in Fig.~\ref{fig:sketch} for \(R_{12} = 2020\,{\rm nm} \approx 2.75\lambda_1\). The QNM expansion from Eq.~\eqref{eq:QNMexp} together with Eq.~\eqref{eq:Bapprox} (\(G^{\rm QNM}\)) yields excellent agreement with the full numerical Green's function calculation (\(G^{\rm full}\)). Using Eq.~\eqref{eq:Bpole} instead of Eq.~\eqref{eq:Bapprox} (\(G^{\rm QNM}_{\rm pole}\)) does not fully include retardation effects. The single-cavity QNM expansion (\(G_1\)) [Eq.~\eqref{eq:singlecavexp}]  does not match the full numerical results, confirming the coupling of the QNMs. The coupling is normalized to  \(g^{\rm back}_{ba}(\omega_1)\) (i.e., without the dimers).}
    \label{fig:Gcomp_d2000}
\end{figure}

For comparison, we also show in Fig.~\ref{fig:Gcomp_d2000} a single-cavity expansion of the Green's function [cf.~Eq.~\eqref{eq:singlegreenoutside}, \(1\leftrightarrow 2 \)]
\begin{align}\label{eq:singlecavexp}
    g_{ba}(\omega)\big|^{\rm single}\approx \frac{1}{\epsilon_0}\mathbf{d}_b\cdot\mathbf{G}_1(\mathbf{r}_b,\mathbf{r}_a,\omega)\cdot\mathbf{d}_a,
\end{align}
where we assume that the electric field of dipole \(a\) (located in cavity \(\mathcal{V}_1\)) can be approximated in terms of the QNMs of cavity \(1\) only. Evidently, this expansion 
{\it significantly underestimates} the strength of the dipole coupling and fails to capture the specific shape which is dominated by the {\it inter-cavity} QNM coupling,
which is contained in the expansion from Eq.~\eqref{eq:QNMexp} but not in Eq.~\eqref{eq:singlecavexp}.

In Fig.~\ref{fig:Gcomp_d740}, we show the coupling \(g_{ba}(\omega)\) for a shorter dimer separation of \(R_{21} = 760\,{\rm nm}\), which is close to the QNM resonance wavelengths. The QNM expansion from Eq.~\eqref{eq:QNMexp} together with Eq.~\eqref{eq:Bapprox} [where \red{\(N_{21} = (5.3797 + i0.4523)\cdot 10^{-7}\,{\rm nm}\)}] again shows excellent agreement with full numerical calculations. Including the phase in the pole approximation [Eq.~\eqref{eq:Bpole}] yields better agreement with the full solution than for \(R_{21}=2020\,{\rm nm}\), since retardation effects are weaker for shorter separations. The single-cavity expansion 
[Eq.~\eqref{eq:singlecavexp}] again fails to capture the full coupling, both quantitatively and qualitatively.

\begin{figure}
    \centering
    \includegraphics[width=0.9\linewidth]{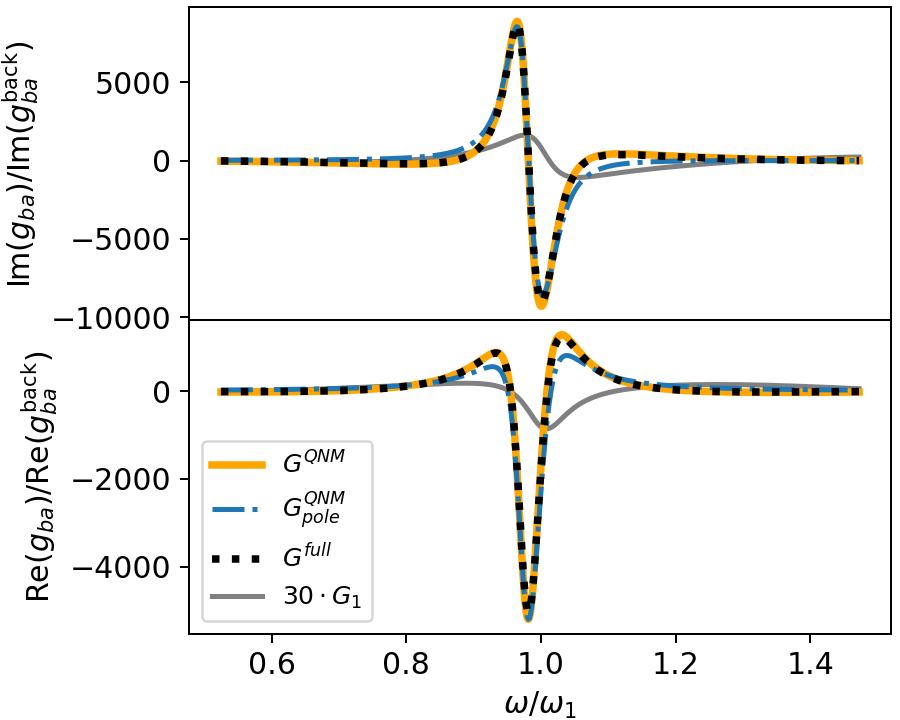}
    \caption{Same as in Fig.~\ref{fig:Gcomp_d2000}, but with \(R_{12} = 760\,{\rm nm} \approx 1.04\lambda_1\).
    Equation~\eqref{eq:QNMexp} together with Eq.~\eqref{eq:Bapprox} (\(G^{\rm QNM}\)) agrees excellently with the full numerical solution (\(G^{\rm full}\)). Equation~\eqref{eq:Bpole} (\(G^{\rm QNM}_{\rm pole}\)) yields better agreement than for larger separations, while the single-cavity QNM expansion (\(G_1\)) from Eq.~\eqref{eq:singlecavexp} fails to capture the shape and magnitude of the coupling.}
    \label{fig:Gcomp_d740}
\end{figure}

\textit{Conclusions and 
discussion.---}We have presented a multi-cavity expansion of the electromagnetic Green's function using single-cavity QNMs. Starting from the single-cavity QNM Green's function, the scheme iteratively adds more cavities via a set of scattering equations where the Green's function obtained from each step serves as input in the next step. Thus, the full scattered Green's function for an arbitrary number of cavities is easily obtained from a finite number of iterations.

Fixing a well known problem for coupled QNMs over large distances, our framework permits an accurate few-mode approximation by using regularization and including retardation effects, making it suitable for quantum dynamics applications where the Hilbert space scales exponentially with the number of modes. Multi-cavity scattering terms naturally decompose into products of two-cavity scattering processes, avoiding nested integrals. 

Comparing rigorous numerical solutions of the full 3D Maxwell equations to the QNM expansion, we demonstrated the quantitative accuracy of the approach for two coupled metal dimers serving as QNM cavities at two different distances between the dimers. 

The scheme presented here may help in the study and design of novel quantum devices by simplifying their numerical simulation and providing an intuitive framework for scattering between open resonators with finite retardation delays.

R.M.F. and M.R. acknowledge support from the Deutsche Forschungsgemeinschaft (Project number 525575745).
J.R. and S.H. acknowledge funding from Queen's University, Canada, 
the Canadian Foundation for Innovation (CFI), 
the Natural Sciences and Engineering Research Council of Canada (NSERC), and CMC Microsystems for the provision of COMSOL Multiphysics;  
S.H. also thanks the 
Alexander von Humboldt Foundation 
for support through a Humboldt Award.

\bibliography{thebiblio}

\clearpage
\appendix
\setcounter{page}{1}
\setcounter{figure}{0}
\setcounter{equation}{0}
\counterwithout{equation}{section}
\renewcommand{\theequation}{S\arabic{equation}}
\renewcommand{\thefigure}{S\arabic{figure}}
\renewcommand{\thepage}{S\arabic{page}}
\renewcommand{\thesection}{S\arabic{section}}
\renewcommand{\thesubsection}{\Alph{subsection}}
\renewcommand{\appendixname}{Supplemental Material}
\title{Supplemental material for ``Green's function expansion for multiple coupled optical resonators with finite retardation using quasinormal modes''}



\maketitle
\onecolumngrid
\section{Numerical details}
\subsection{Model set-up}\label{appsec:setup}
We consider two metallic dimers in vacuum (\(\epsilon_{\rm back} = 1\)). Dimer 1 consists of two identical cylindrical nanorods of length \(L_1 = 80\, {\rm nm}\) and base radius \(r_1 = 10\,{\rm nm}\). The gap distance between the nanorods along the \(z\)-axis is \(d_1 = 10\,{\rm nm}\). Dimer 2 consists of two identical nanorods with \(L_2 = 90\,{\rm nm}\), \(r_2 = 10\,{\rm nm}\), and \(d_2 = 10\,{\rm nm}\).
The metal dimers are modeled using Drude permittivities \begin{align}\label{appeq:Drude}
    \epsilon_j(\omega)=1-\frac{\omega_{\rm p}^{2}}{\omega^{2}+i\omega\gamma_{{\rm p}j}},
 \end{align}
where \(j=1,2\), and \(\hbar\omega_{\rm p} = 8.2934\,{\rm eV}\) is the plasma frequency, while  \(\hbar\gamma_{\rm p1}=0.0928\,{\rm eV}\) and  \(\hbar\gamma_{\rm p2}=0.3\hbar\gamma_{\rm p1}\). 

The single-cavity QNMs are calculated using an inverse Green's function approach in complex frequency space \cite{bai_efficient_2013-1,ren2020near}, with the aid of COMSOL Multiphysics (a finite element solver). The computational model is a cylinder with a diameter of $3$~$\mu$m and a length of $5.2$~$\mu$m, resulting in a total volume of around $37$ $\mu$m$^{3}$, which contains perfectly matched layers (PMLs) with a thickness of $600~$nm to minimize boundary reflections efficiently. 
High-resolution mesh settings are used, where the maximum mesh element size around the dipole point, inside and outside the metallic nanorods, are $0.1$ nm, $2$ nm, and $80~$nm, respectively.

The QNM of dimer \(1\) has the eigenfrequency \(\hbar\tilde{\omega}_1=\hbar\omega_1-i\hbar\gamma_1=(1.6904-0.0652i)\,{\rm eV}\), corresponding to a quality factor of \(Q_1=\omega_1/(2\gamma_1)\approx13.0\), and a wavelength of \(\lambda_1 \approx 733.46\,{\rm nm}\). For the QNM of dimer \(2\), we obtain \(\hbar\tilde{\omega}_2=\hbar\omega_2-i\hbar\gamma_2=(1.6482-0.0388i)\,{\rm eV}\), and thus \(Q_2\approx21.2\) and \(\lambda_2 \approx 752.24\,{\rm nm}\).

\subsection{Numerical Greens function}
The Green's function \(\mathbf{G}({\mathbf{r}_{a}},{\mathbf{r}_{b}},\omega)\) between two dipoles \(\mathbf{r}_a\in\mathcal{V}_1\), \(\mathbf{r}_b\in\mathcal{V}_2\) in the gap centers of two metal dimers, \(1\) and \(2\), can be obtained numerically from
\begin{align}
\mathbf{E}({\mathbf{r}_{a}},\omega)&=\frac{\mathbf{d}_{b}\cdot{\mathbf{G}({\mathbf{r}_{b}},{\mathbf{r}_{a}},\omega)}}{\epsilon_{0}},
\end{align}

i.e., we retrieve the Green's functions $\mathbf{G}({\mathbf{r}_{b}},{\mathbf{r}_{a}},\omega)$ from the electric field $\mathbf{E}({\mathbf{r}_{a}},\omega)$ by placing a dipole with real dipole moment $\mathbf{d}_b$ at $\mathbf{r}_{b}$ (gap center of the dimer 2). Similarly, we can get $\mathbf{G}(\mathbf{r}_{a},\mathbf{r}_{b},\omega)$ through $\mathbf{E}(\mathbf{r}_{b},\omega)$ by placing a dipole \(\mathbf{d}_a\) at \(\mathbf{r}_a\) (gap center of dimer 1).

\subsection{Numerical calculation of \(N_{21}\)}
As shown in the main document, the QNM expansion of the Green's function between the two dipoles \(\mathbf{r}_a\in\mathcal{V}_1\), \(\mathbf{r}_b\in\mathcal{V}_2\) reads,
\begin{align}\label{appeq:Gexp}
    &\mathbf{G}(\mathbf{r}_b,\mathbf{r}_a,\omega)\big|_{ \mathbf{r}_b \in \mathcal{V}_2,\mathbf{r}_a\in \mathcal{V}_1}=  B_{21}(\omega)\qnm{f}{2}(\mathbf{r}_b)\qnm{f}{1}(\mathbf{r}_a),
\end{align}
where
\begin{align}\label{appeq:Bsurface}
    B_{21}(\omega) = \frac{c^2}{\omega^2}A_2(\omega)A_1(\omega)\oint_{\mathcal{S}_1}\mathrm{d}A_s \Big\{\Big[\nabla_s\times\qnm{F}{2}(\mathbf{s},\omega)\Big]\cdot\Big[\opvec{n}_s\times\qnm{f}{1}(\mathbf{s})\Big]-\Big[\opvec{n}_s\times\qnm{F}{2}(\mathbf{s},\omega)\Big]\cdot\Big[\nabla_s\times\qnm{f}{1}(\mathbf{s})\Big]\Big\},
\end{align}
and \(B_{21}(\omega)=B_{12}(\omega)\) is symmetric in the cavity indices.

As discussed in the main document, we decompose \(\qnm{F}{2}(\mathbf{r},\omega)\big|_{\mathbf{r}\in\mathcal{V}_1} = \qnm{F}{2}'(\mathbf{r},\omega)\mathrm{e}^{i\omega R_{21}/c}\) and then apply a pole approximation \(\omega\to\omega_2\) \cite{ren2020near} (where \(\omega_2\) is the central mode frequency of QNM 2) to the slowly-varying envelope function \(\qnm{F}{2}'(\mathbf{r},\omega)\), so that
\begin{align}\label{appeq:Bapprox}
    B_{21}(\omega) \approx N_{21} \frac{c^2}{\omega^2}A_2(\omega)A_1(\omega)\mathrm{e}^{i\omega R_{21}/c},
\end{align}
with
\begin{align}\label{appeq:N21}
    N_{21} &= \oint_{\mathcal{S}_1}\mathrm{d}A_s \Big\{\Big[\nabla_s\times\qnm{F}{2}'(\mathbf{s},\omega_2)\Big]\cdot\Big[\opvec{n}_s\times\qnm{f}{1}(\mathbf{s})\Big]-\Big[\opvec{n}_s\times\qnm{F}{2}'(\mathbf{s},\omega_2)\Big]\cdot\Big[\nabla_s\times\qnm{f}{1}(\mathbf{s})\Big]\Big\}.
\end{align}

\(N_{21}\) can be rewritten as 
\begin{align}\label{appeq:N21_calcu}
    N_{21} &= \oint_{\mathcal{S}_1}\mathrm{d}A_s \Big\{\Big[-i \mu_0 \omega_2\qnm{H}{2}'(\mathbf{s},\omega_2)\cdot\qnm{M}{1}(\mathbf{s})\Big]+\Big[i\mu_0 \tilde{\omega}_1\qnm{F}{2}'(\mathbf{s},\omega_2)\cdot\qnm{J}{1}(\mathbf{s})\Big]\Big\},
\end{align}
by using (cf.~Ref.~\onlinecite{ren2020near})
\begin{align}
    &\qnm{H}{2}'(\mathbf{s},\omega_2) = \nabla_s\times\qnm{F}{2}'(\mathbf{s},\omega_2)/(i\mu_0 \omega_2),\nonumber\\
    &\qnm{h}{1}(\mathbf{s}) = \nabla_s\times\qnm{f}{1}(\mathbf{s})/(i\mu_0\tilde{\omega}_1),\nonumber\\
    &\qnm{J}{1}(\mathbf{s})=\opvec{n}_s\times\qnm{h}{1}(\mathbf{s}),\nonumber\\
    &\qnm{M}{1}(\mathbf{s}) = -\opvec{n}_s\times\qnm{f}{1}(\mathbf{s}).
\end{align}

To numerically calculate $N_{21}$, a closed cuboid surface which is $50~$nm away from the dimer 1 is selected as the near-field surface $\mathcal{S}_1$.
A relatively small grid size, $\rm (0.5~ nm)^2$, of the surface integral for $\qnm{M}{1}(\mathbf{s})$ and $\qnm{J}{1}(\mathbf{s})$ is employed as they change fast in the near-field region. For $\qnm{F}{2}'(\mathbf{s},\omega_2)$
and $\qnm{H}{2}'(\mathbf{s},\omega_2)$, we use a relative large grid element, $\rm (5~ nm)^2$, as they change slowly in the near field surface of dimer 1 (assuming dimer 2 is far apart from dimer 1).

\red{
\section{Mathematical details}
\subsection{Derivation of the Dyson equation}\label{appsec:dysonderiv}
The Green's function for \(n\) cavities solves the Helmholtz equation
\begin{align} \label{appeq:greenhelm}
    \Big[\nabla\times\nabla\times-\frac{\omega^2}{c^2}\epsilon^{(n)}(\mathbf{r},\omega)\Big]\mathbf{G}^{(n)}(\mathbf{r},\mathbf{r}',\omega)=\frac{\omega^2}{c^2}\mathbb{1}\delta(\mathbf{r}-\mathbf{r}'),
\end{align}
under outgoing boundary conditions such as the Silver-Müller radiation condition for cavities in three-dimensional homogeneous background media \cite{muller1948grundzuge, silver1984microwave}. 

Using the Helmholtz equation~\eqref{appeq:greenhelm} with \(\epsilon^{(n)}(\mathbf{r},\omega)\) for the Green's function \(\mathbf{G}^{(n)}(\mathbf{r},\mathbf{r}',\omega)\) and with \(\epsilon^{(n-1)}(\mathbf{r},\omega)\) for \(\mathbf{G}^{(n-1)}(\mathbf{r},\mathbf{r}',\omega)\), we find (omitting the \(\omega\)-dependence of the Green's functions for brevity)
\begin{align*}
    \mathbf{G}^{(n)}(\mathbf{r},\mathbf{r}')-\mathbf{G}^{(n-1)}(\mathbf{r},\mathbf{r}') &= \int\mathrm{d}^3s \, \Big[\delta(\mathbf{r}-\mathbf{s})\mathbf{G}^{(n)}(\mathbf{s},\mathbf{r}')-\mathbf{G}^{(n-1)}(\mathbf{r},\mathbf{s})\delta(\mathbf{r}'-\mathbf{s})\Big]\nonumber\\
    &=\frac{c^2}{\omega^2}\int\mathrm{d}^3s\Big\{\Big[\nabla\times\nabla\times\mathbf{G}^{(n-1)}(\mathbf{s},\mathbf{r})\Big]^T \cdot\mathbf{G}^{(n)}(\mathbf{s},\mathbf{r}')-\mathbf{G}^{(n-1)}(\mathbf{r},\mathbf{s})\cdot\Big[\nabla\times\nabla\times\mathbf{G}^{(n)}(\mathbf{s},\mathbf{r}')\Big] \Big\}\nonumber\\
    &\quad+\int\mathrm{d}^3s\, \mathbf{G}^{(n-1)}(\mathbf{r},\mathbf{s})\cdot\mathbf{G}^{(n)}(\mathbf{s},\mathbf{r}')\Big[\epsilon^{(n)}(\mathbf{s},\omega)-\epsilon^{(n-1)}(\mathbf{s},\omega)\Big].
\end{align*}
The first term on the right-hand side can be turned into a surface integral over a far-field surface, which vanishes since there is no scattering of the outgoing waves in the far field \cite{ge2014quasinormal, franke2020fluctuation}. In the second term, the expression in the parentheses vanishes everywhere except inside the \(n\)-th cavity, where \(\epsilon^{(n-1)}(\mathbf{s},\omega)\big|_{\mathbf{s}\in\mathcal{V}_n} = \epsilon_{\rm back}(\mathbf{s},\omega)\), yielding the Dyson equation from the main text
\begin{align}\label{appeq:dyson}
    \mathbf{G}^{(n)}(\mathbf{r},\mathbf{r}',\omega) = \mathbf{G}^{(n-1)}(\mathbf{r},\mathbf{r}',\omega)+\int_{\mathcal{V}_n}\mathrm{d}^3s \Delta\epsilon(\mathbf{s},\omega)\mathbf{G}^{(n-1)}(\mathbf{r},\mathbf{s},\omega)\cdot\mathbf{G}^{(n)}(\mathbf{s},\mathbf{r}',\omega),
\end{align}
where \(\Delta\epsilon(\mathbf{s},\omega) = \epsilon(\mathbf{s},\omega)-\epsilon_{\rm back}(\mathbf{s},\omega)\).}
\red{
\subsection{Derivation of the surface integral representation}\label{appsec:surfacederiv}
For the surface integral representation, we again use the Helmholtz equation on the Dyson equation~\eqref{appeq:dyson} to find,
\begin{align*}
    \mathbf{G}^{(n)}(\mathbf{r},\mathbf{r}') &= \mathbf{G}^{(n-1)}(\mathbf{r},\mathbf{r}') +\int_{\mathcal{V}_n}\mathrm{d}^3s \mathbf{G}^{(n-1)}(\mathbf{r},\mathbf{s})\cdot\mathbf{G}^{(n)}(\mathbf{s},\mathbf{r}')\Big[\epsilon^{(n)}(\mathbf{s},\omega)-\epsilon^{(n-1)}(\mathbf{s},\omega)\Big]\nonumber\\
    &=\mathbf{G}^{(n-1)}(\mathbf{r},\mathbf{r}')+\int_{\mathcal{V}_n}\mathrm{d}^3s \Big[\delta(\mathbf{r}-\mathbf{s})\mathbf{G}^{(n)}(\mathbf{s},\mathbf{r}')-\mathbf{G}^{(n-1)}(\mathbf{r},\mathbf{s})\delta(\mathbf{s}-\mathbf{r}')\Big]\nonumber\\
    &\quad-\frac{c^2}{\omega^2}\int_{\mathcal{V}_n}\mathrm{d}^3s \Big\{\Big[\nabla_s\times\nabla_s\times\mathbf{G}^{(n-1)}(\mathbf{s},\mathbf{r})\Big]^T\cdot\mathbf{G}^{(n)}(\mathbf{s},\mathbf{r}')-\mathbf{G}^{(n-1)}(\mathbf{r},\mathbf{s})\cdot\Big[\nabla_s\times\nabla_s\times\mathbf{G}^{(n)}(\mathbf{s},\mathbf{r}')\Big]\Big\}.
\end{align*}

The last integral on the right-hand side is turned into an integral over the cavity surface \(\mathcal{S}_n\) using Green's second identity. 
For \(\mathbf{r}\notin \mathcal{V}_n\), the first delta function on the right hand side vanishes (for cases with \(\mathbf{r}\in\mathcal{V}_n,\mathbf{r}'\notin\mathcal{V}_n\), the transpose of Eq.~\eqref{eq:dyson} is used instead, so that the derivation discussed here still holds), yielding the surface integral representation from the main text:
\begin{align}\label{appeq:dysonsurface}
    \mathbf{G}^{(n)}(\mathbf{r},\mathbf{r}')\Big|_{\mathbf{r}\notin\mathcal{V}_n}= \chi_{\overline{\mathcal{V}}_n}(\mathbf{r}')\mathbf{G}^{(n-1)}(\mathbf{r},\mathbf{r}')+\frac{c^2}{\omega^2}\oint_{\mathcal{S}_n}\mathrm{d}A_s &\Big\{\Big[\nabla_s\times\mathbf{G}^{(n-1)}(\mathbf{s},\mathbf{r})\Big]^T\cdot\Big[\opvec{n}_s\times\mathbf{G}^{(n)}(\mathbf{s},\mathbf{r}')\Big]\nonumber\\
    &-\Big[\opvec{n}_s\times\mathbf{G}^{(n-1)}(\mathbf{s},\mathbf{r})\Big]^T\cdot\Big[\nabla_s\times\mathbf{G}^{(n)}(\mathbf{s},\mathbf{r}')\Big]\Big\}.
\end{align}
}
\red{
\subsection{Perturbative treatment of the Dyson equation}\label{appsec:perturbative}
In the main text, we evaluate the Dyson equation~\eqref{appeq:dyson} using the assumption that the multi-cavity Green's function inside one of the cavities can be expanded in terms of the QNMs of that cavity alone. This assumption holds very well for the example of two coupled metal dimers with significant retardation. In general, a perturbative treatment of Eq.~\eqref{appeq:dyson} is also possible, but needs to be performed carefully, since \(\Delta\epsilon(\mathbf{r},\omega)\) is usually not a small perturbation. Instead, a perturbative series can be expanded in orders of the \textit{intercavity scattering}, which is often small compared to the intracavity scattering. As an example, we consider the case of two coupled cavities with significant retardation. The extension to more cavities is straightforward. 

The Dyson series from Eq.~\eqref{appeq:dyson} for the Green's function for two cavities \(\mathbf{G}\equiv \mathbf{G}^{(2)}\) reads, 
\begin{align}\label{appeq:twodyson}
    \mathbf{G}(\mathbf{r},\mathbf{r}') =  \mathbf{G}_1(\mathbf{r},\mathbf{r}')+\int_{\mathcal{V}_2}\mathrm{d}^3s \Delta\epsilon(\mathbf{s}) \mathbf{G}_1(\mathbf{r},\mathbf{s})\cdot \mathbf{G}(\mathbf{s},\mathbf{r}'),
\end{align}
where we omitted the \(\omega\)-dependence for brevity and used the single-cavity Green's function of cavity 1 alone, \(\mathbf{G}_1\), as a basis for the expansion. Note that an equivalent expansion using \(\mathbf{G}_2\) is possible by switching \(1\leftrightarrow 2\) in Eq.~\eqref{appeq:twodyson}. A transposed version of Eq.~\eqref{appeq:twodyson} can also be derived following the steps laid out in Sec.~\ref{appsec:dysonderiv}. Hence, an alternative formulation of Eq.~\eqref{appeq:twodyson} reads,
\begin{align}\label{appeq:alttwodyson}
     \mathbf{G}(\mathbf{r},\mathbf{r}') =  \mathbf{G}_2(\mathbf{r},\mathbf{r}')+\int_{\mathcal{V}_1}\mathrm{d}^3s \Delta\epsilon(\mathbf{s}) \mathbf{G}(\mathbf{r},\mathbf{s})\cdot \mathbf{G}_2(\mathbf{s},\mathbf{r}').
\end{align}

Which version of the Dyson equation is best employed depends on the positions arguments. For the case \(\mathbf{r}\in\mathcal{V}_1,\mathbf{r}'\in\mathcal{V}_2\), we use a combination of Eqs.~\eqref{appeq:twodyson} and \eqref{appeq:alttwodyson}, to obtain
\begin{align*}
    \mathbf{G}(\mathbf{r},\mathbf{r}')\big|_{\mathbf{r}\in\mathcal{V}_1,\mathbf{r}'\in\mathcal{V}_2} &=  \frac{\mathbf{G}_1(\mathbf{r},\mathbf{r}')+\mathbf{G}_2(\mathbf{r},\mathbf{r}')}{2}+\frac{1}{2}\int_{\mathcal{V}_2}\mathrm{d}^3s \Delta\epsilon(\mathbf{s}) \mathbf{G}_1(\mathbf{r},\mathbf{s})\cdot \mathbf{G}(\mathbf{s},\mathbf{r}')+\frac{1}{2}\int_{\mathcal{V}_1}\mathrm{d}^3s \Delta\epsilon(\mathbf{s}) \mathbf{G}(\mathbf{r},\mathbf{s})\cdot \mathbf{G}_2(\mathbf{s},\mathbf{r}').
\end{align*}

Here, the integrals on the right-hand side contain intercavity transfer terms, where the field from one cavity propagates to the other cavity. At the same time, the full Green's functions that appear on the right-hand side are located inside one cavity (\(\mathbf{s},\mathbf{r}'\in\mathcal{V}_2\) for the \(\mathcal{V}_2\)-integral, and \(\mathbf{r},\mathbf{s}\in\mathcal{V}_1\) for the \(\mathcal{V}_1\)-integral). In the main text, we apply the termination condition at this point. For a perturbative expansion to first order in the intercavity terms, we reinsert Eq.~\eqref{appeq:twodyson} into the \(\mathcal{V}_2\) integral and Eq.~\eqref{appeq:alttwodyson} into the \(\mathcal{V}_1\)-integral, to obtain
\begin{align}\label{appeq:dysonpert}
    &\mathbf{G}(\mathbf{r},\mathbf{r}')\big|_{\mathbf{r}\in\mathcal{V}_1,\mathbf{r}'\in\mathcal{V}_2}\nonumber\\
    &=  \frac{\mathbf{G}_1(\mathbf{r},\mathbf{r}')+\mathbf{G}_2(\mathbf{r},\mathbf{r}')}{2}+\frac{1}{2}\int_{\mathcal{V}_2}\mathrm{d}^3s \Delta\epsilon(\mathbf{s}) \mathbf{G}_1(\mathbf{r},\mathbf{s})\cdot \mathbf{G}_2(\mathbf{s},\mathbf{r}')+\frac{1}{2}\int_{\mathcal{V}_1}\mathrm{d}^3s \Delta\epsilon(\mathbf{s}) \mathbf{G}_1(\mathbf{r},\mathbf{s})\cdot \mathbf{G}_2(\mathbf{s},\mathbf{r}')+\mathcal{O}[(\Delta\epsilon^{\rm inter})^2],
\end{align}
where \((\Delta\epsilon^{\rm inter})^2\) refers to second-order intercavity scattering terms. Under the assumption that the intercavity scattering is small compared to intracavity scattering, we neglect the terms \(\mathcal{O}[(\Delta\epsilon^{\rm inter})^2]\), so that Eq.~\eqref{appeq:dysonpert} contains only the \textit{single-cavity} Green's functions \(\mathbf{G}_1,\mathbf{G}_2\), for which QNM expansions are known. Since the single-cavity Green's function factorizes into dyadic products of the QNMs, multi-cavity scattering nested integrals decompose into products of two-cavity scattering integrals for the perturbative expansion.

Perturbative series for more than two cavities and higher orders of the intercavity scattering may be constructed similarly, depending on the specific structure. From Eq.~\eqref{appeq:dysonpert}, a surface-integral representation may be obtained, following the steps from Sec.~\ref{appsec:surfacederiv}. We note that, for two cavities, the termination condition from the main text yields the same result as a first-order perturbative expansion in the intercavity scattering. 
}
\red{
\subsection{Three-cavity Green's function}
From Eq.~\eqref{appeq:dysonsurface}, we obtain, for three coupled cavities (omitting the \(\omega\)-dependence),
\begin{align}\label{appeq:threecavdyson}
    &\mathbf{G}^{(3)}(\mathbf{r},\mathbf{r}')\big|_{\mathbf{r}\in \mathcal{V}_1, \mathbf{r}' \in \mathcal{V}_3} \nonumber\\
    &\quad= \frac{c^2}{\omega^2}\oint_{\mathcal{S}_3}\mathrm{d}A_s \Big\{\Big[\nabla_s\times\mathbf{G}^{(2)}(\mathbf{s},\mathbf{r})\Big]^T\cdot\Big[\opvec{n}_s\times\mathbf{G}^{(3)}(\mathbf{s},\mathbf{r}',)\Big]-\Big[\opvec{n}_s\times\mathbf{G}^{(2)}(\mathbf{s},\mathbf{r})\Big]^T\cdot\Big[\nabla_s\times\mathbf{G}^{(3)}(\mathbf{s},\mathbf{r}')\Big]\Big\}.
\end{align}

The two-cavity Green's function \(\mathbf{G}^{(2)}(\mathbf{s},\mathbf{r},\omega)\) is obtained in a way similar to what is shown in the main text, but for \(\mathbf{s}\) outside the cavities \(1\) and \(2\), and reads, for one dominant QNM per cavity,
\begin{align}
    \mathbf{G}^{(2)}(\mathbf{s},\mathbf{r},\omega)\big|_{\mathbf{s}\notin\mathcal{V}_1\cup\mathcal{V}_2,\mathbf{r}\in\mathcal{V}_1} = A_1(\omega)\qnm{F}{1}(\mathbf{s},\omega)\qnm{f}{1}(\mathbf{r})+B_{21}(\omega)\qnm{F}{2}(\mathbf{s},\omega)\qnm{f}{1}(\mathbf{r}),
\end{align}
while \(\mathbf{G}^{(3)}(\mathbf{s},\mathbf{r}',\omega)\big|_{\mathbf{s},\mathbf{r}'\in\mathcal{V}_3}\) fulfills the condition \(\mathbf{G}^{(3)}(\mathbf{s},\mathbf{r}',\omega)\big|_{\mathbf{s},\mathbf{r}'\in\mathcal{V}_3} = \mathbf{G}_3(\mathbf{s},\mathbf{r},\omega)\) from the main text. 

Inserting these results into Eq.~\eqref{appeq:threecavdyson}, we obtain
\begin{align*}
    \mathbf{G}^{(3)}(\mathbf{r},\mathbf{r}',\omega)\big|_{\mathbf{r}\in \mathcal{V}_1, \mathbf{r}' \in \mathcal{V}_3} = [B_{13}(\omega)+C_{123}(\omega)]\qnm{f}{1}(\mathbf{r})\qnm{f}{3}(\mathbf{r}'),
\end{align*}
where the first term in the parentheses gives the direct scattering between cavities 1 and 3, while the second term gives the scattering mediated by cavity 2.
The three-cavity coupling element reads,
\begin{align}
    C_{123}(\omega)& = B_{21}(\omega)A_3(\omega)\frac{c^2}{\omega^2}\oint_{\mathcal{S}_3}\mathrm{d}A_s \Big\{\Big[\nabla_s\times\qnm{F}{2}(\mathbf{s},\omega)\Big]\cdot\Big[\opvec{n}_s\times\qnm{f}{3}(\mathbf{s})\Big]-\Big[\opvec{n}_s\times\qnm{F}{2}(\mathbf{s},\omega)\Big]\cdot\Big[\nabla_s\times\qnm{f}{3}(\mathbf{s})\Big]\Big\}\nonumber\\
    &= B_{21}(\omega)B_{23}(\omega)/A_2(\omega).
\end{align}
Here, the power of the QNM expansion becomes apparent: The multi-cavity scattering process decomposes into a product of independently calculable two-cavity scattering processes, so that no costly convoluted integrals have to be calculated, and the number of scattering integrals \(N_{ij}\) that have to be calculated scales quadratically with the number of cavities within a single-mode approximation. Note that, while we employed Eq.~(8) here for brevity, this decomposition of multi-cavity scattering nested integrals is a consequence of the shape of the single-cavity Green's function, where the Green's function factorizes into the dyadic product of two QNMs, and also holds for perturbative approaches (cf.~Sec.~\ref{appsec:perturbative}).
}

\section{Additional results}
\subsection{Comparison between \(N_{21}\) and \(N_{12}\)}
While \(B_{21}(\omega) = B_{12}(\omega)\) is symmetric in the indices, \(N_{21}\neq N_{12}\) due to the approximations we employed (slow-rotating envelope approximation \cite{fuchs2024quantization} and pole approximation \cite{ren2020near}). In the case of the coupled dimers with a center-to-center distance of \(R_{21} = 2020\,{\rm nm}\), we find \red{\(N_{21} = (2.0773-i0.0657)\cdot 10^{-7}\,{\rm nm}^{-2}\)} and \red{\(N_{12} = (2.0608-i0.0610)\cdot 10^{-7}\,{\rm nm}^{-2}\)}\red{, which differ by less than one percent.}

\begin{figure}
    \centering
    \includegraphics[width=0.5\linewidth]{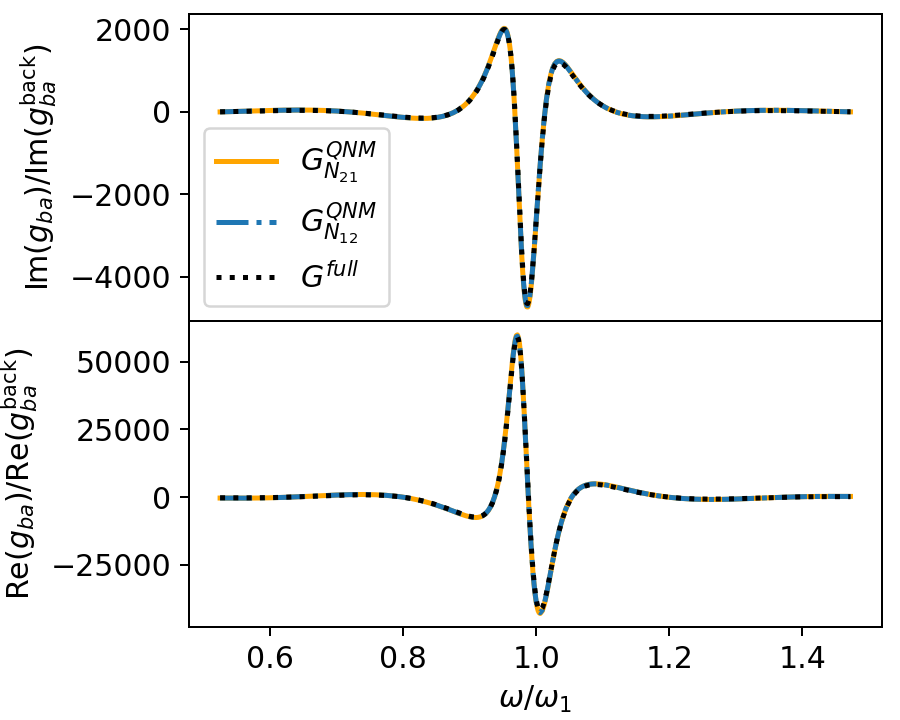}
    \caption{Coupling \(g_{ba}(\omega)\) [cf.~Eq.~\eqref{appeq:gba}] between two  dipoles in the dimer gaps of two metal dimers, for a distance of \(R_{21} = 2020\,{\rm nm}\). We compare the QNM expansion of the Green's function from Eq.~\eqref{appeq:Gexp} together with Eq.~\eqref{appeq:Bapprox} for \(N_{21}\) and \(N_{12}\) to the full numerical Green's function. \red{Both values agree} qualitatively and quantitatively with the full results. The coupling is normalized to the coupling \(g_{ba}^{\rm back}(\omega_1)\) without the presence of the dimers at the resonance frequency of QNM 1.}
    \label{fig:Gcomp_N12vN21}
\end{figure}

In Fig.~\ref{fig:Gcomp_N12vN21}, we show the coupling 
\begin{align}\label{appeq:gba}
    g_{ba}(\omega) = \frac{1}{\epsilon_0}\mathbf{d}_b\cdot\mathbf{G}(\mathbf{r}_b,\mathbf{r}_a,\omega)\cdot\mathbf{d}_a
\end{align}
between two dipoles in the gap centers of the metal dimers. We compare the full numerical Green's function to the expansion from Eq.~\eqref{appeq:Gexp} together with Eq.~\eqref{appeq:Bapprox} for both \(N_{12}\) and \(N_{21}\). \red{We observe excellent agreement with the full solution for both values, without visible difference between \(N_{12}\) and \(N_{21}\).}

\subsection{Results for additional distances}
\begin{figure}
    \centering
    \includegraphics[width=0.5\linewidth]{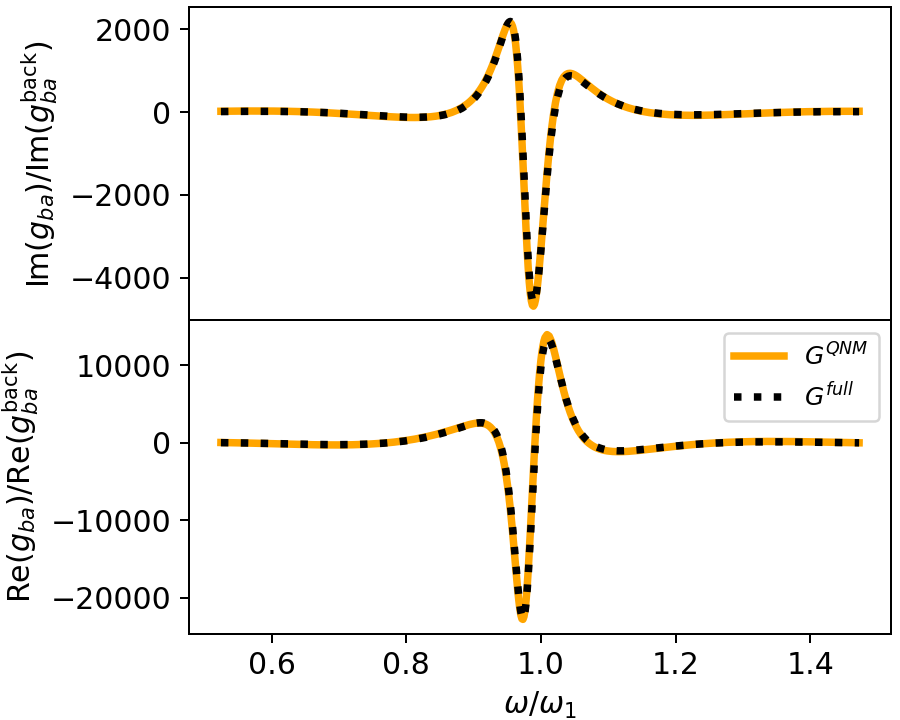}
    \caption{Coupling \(g_{ba}(\omega)\) [cf.~Eq.~\eqref{appeq:gba}] between two dipoles in the gaps of metal dimers serving as QNM cavities for a center-to-center distance of \(R_{21} = 1620\,{\rm nm} \approx 2.21\lambda_1\) between the dimers. The QNM expansion from Eq.~\eqref{appeq:Gexp} together with Eq.~\eqref{appeq:Bapprox} yields excellent agreement with the full numerical Green's function expansion. The coupling is normalized to the coupling \(g_{ba}^{\rm back}(\omega_1)\) without the presence of the dimers at the resonance frequency of QNM 1.}
    \label{fig:Gcomp_d1600}
\end{figure}

In Fig.~\ref{fig:Gcomp_d1600}, we show the coupling \(g_{ba}(\omega)\) [cf.~Eq.~\eqref{appeq:gba}] between the two  dipoles in the dimer gaps for a center-to-center distance of \(R_{21} = 1620\,{\rm nm}\) between the dimers. Again, the QNM expansion from Eq.~\eqref{appeq:Gexp} together with Eq.~\eqref{eq:Bapprox} [where \red{\(N_{21} = (2.5856-i0.0370)\cdot 10^{-7}\,{\rm nm}^{-2}\)}] agrees very well with the full numerical Green's function. 

\begin{figure}
    \centering
    \includegraphics[width=0.5\linewidth]{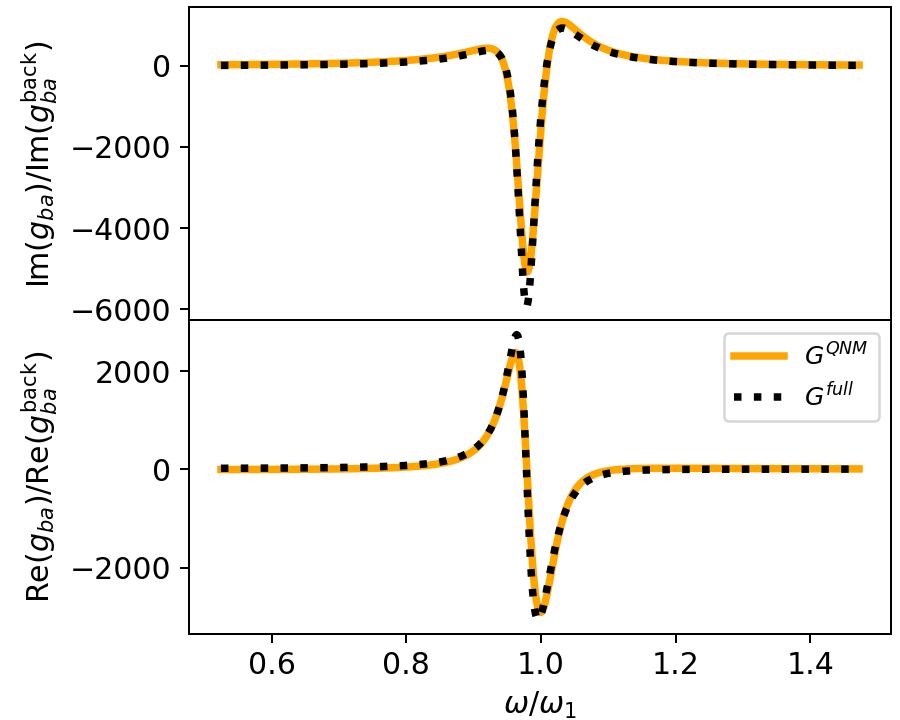}
    \caption{Same as Fig.~\ref{fig:Gcomp_d1600} but for \(R_{21} = 130\,{\rm nm} \approx 0.18\lambda_1\). While the QNM expansion from Eq.~\eqref{appeq:Gexp} together with Eq.~\eqref{appeq:Bapprox} [where \red{\(N_{21}=(13.3588+i13.7713)\cdot 10^{-7}\,{\rm nm}^{-2}\)}] qualitatively captures the shape of the coupling very well, the strength of the coupling is slightly underestimated, especially near the resonance frequency.}
    \label{fig:Gcomp_d110}
\end{figure}

In Fig.~\ref{fig:Gcomp_d110}, we show the same coupling for a center-to-center distance between the dimers of \(R_{21} = 130\,{\rm nm}\), which is significantly shorter than the QNM wavelengths. The QNM expansion from Eq.~\eqref{appeq:Gexp}, together with Eq.~\eqref{appeq:Bapprox} and \red{\(N_{21}=(13.3588+i13.7713)\cdot 10^{-7}\,{\rm nm}\)} still qualitatively captures the coupling between the dipoles, but underestimates the strength of the coupling, especially near the resonance frequency. This indicates that the assumptions made in the derivation of Eq.~\eqref{appeq:Gexp} hold less well for short separations.

Namely, for a short distance away from the cavity surface, the regularized QNM fields \(\qnm{F}{1/2}(\mathbf{r},\omega)\) give a less accurate estimate of the full field, especially within a few-mode approximation. As a consequence, the coupling \(B_{21}(\omega)\) from Eq.~\eqref{appeq:Bsurface} is less accurate for short separations between the cavities. Additionally, in the derivation of Eq.~\eqref{appeq:Gexp} we used a terminating condition,
\begin{align}\label{appeq:termcond}
    \mathbf{G}(\mathbf{r},\mathbf{r}',\omega)\big|_{\mathbf{r},\mathbf{r}'\in\mathcal{V}_i} = \mathbf{G}_i(\mathbf{r},\mathbf{r}',\omega),
\end{align}
where we assumed that the full Green's function inside (or very close to) a cavity can be expanded in terms of the QNMs of that cavity only. This assumption holds well for large spatial separations between the cavities (and especially in three-dimensional, homogeneous background media), but breaks down for very short separations, where the presence of the other cavities influences the field inside the cavity. 

As Fig.~\ref{fig:Gcomp_d110} shows, the expansion from Eq.~\eqref{appeq:Gexp} still fits the full numerical results relatively well, even for separations that are much shorter than the resonance wavelength. For even shorter separations, where the cavities are in the near-field regime, \red{ a different termination condition from Eq.~\eqref{appeq:termcond} or rigorous higher-order perturbative treatment (cf.~Sec.~\ref{appsec:perturbative}) is likely required.}

Alternatively, other approaches, such as the coupled quasinormal mode theory \cite{PhysRevB.102.045430,ren2021quasinormal, ren2022connecting}, hold very well for coupled resonators with short separations and negligible retardation effects, and thus can be employed in such cases to obtain the full coupled Green's function. 

\red{
\subsection{Identical dimers}
We also calculate the QNM expansion of the Green's function for two identical dimers. We choose the values \(L_1,r_1,d_1\) for both dimers, as well as the same Drude model from Eq.~\eqref{appeq:Drude} with \(\hbar\gamma_{p1}\) for both dimers. Hence, the dimers have dominant QNMs with the same frequency \(\hbar\tilde{\omega}_1 = \hbar\tilde{\omega}_2 = (1.6904-0.0652i)\,{\rm eV}\) (cf.~Sec.~\ref{appsec:setup}). For a center-to-center distance of \(R_{12} = 2020\,{\rm nm}\) between the dimers, we obtain \(N_{21} = (2.0154-i0.0993)\cdot 10^{-7}\,{\rm nm}^{-2}\). The results are shown in Fig.~\ref{fig:identical}, and agree excellently. 

\begin{figure}
    \centering
    \includegraphics[width=0.5\linewidth]{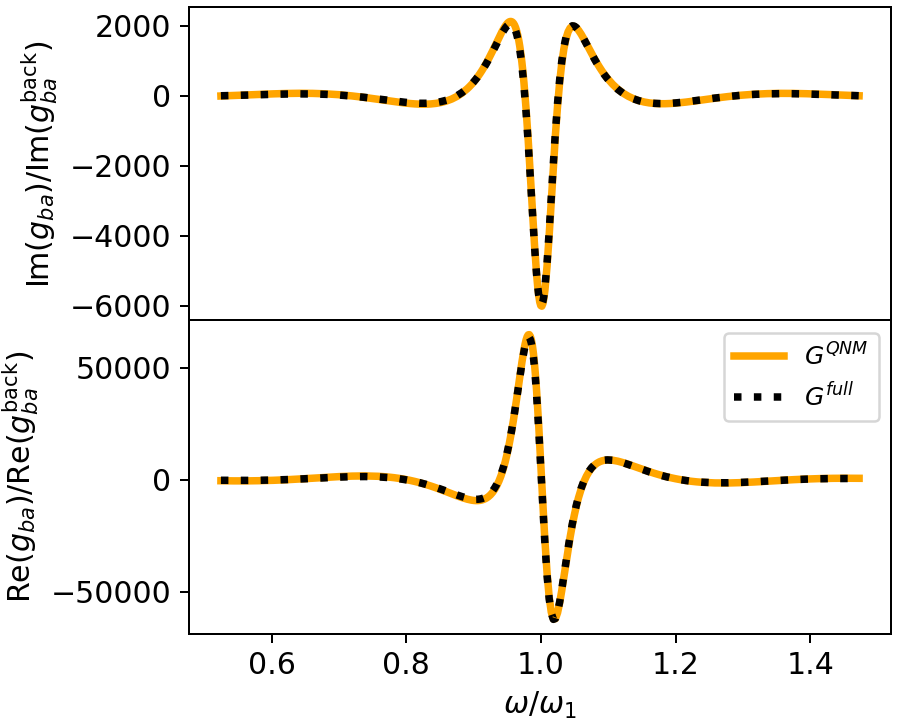}
    \caption{Same as Fig.~\ref{fig:Gcomp_d1600} but for a setup where dimer 2 is identical to dimer 1 (cf.~Sec.~\ref{appsec:setup} and \(R_{12} = 2020\,{\rm nm} = 2.75\lambda_1\). The QNM expansion from Eq.~\eqref{appeq:Gexp} together with Eq.~\eqref{appeq:Bapprox} [where {\(N_{21}=(2.0154-i0.0993)\cdot 10^{-7}\,{\rm nm}^{-2}\)}] agrees excellently with the Green's function result obtained numerically via dipole excitation. }
    \label{fig:identical}
\end{figure}
}
\end{document}